\documentclass[aps,prb,twocolumn,groupedaddress]{revtex4}
\usepackage{graphicx}

\begin{document}


\title{Native defects in hybrid C/BN nanostructures}


\author{J. M. Pruneda}
\email[]{miguel.pruneda@cin2.es}
\affiliation{Centre d'Investigaci\'on en Nanoci\'encia i Nanotecnolog{\'{\i}}a (CSIC-ICN). Campus de la UAB, E-08193 Bellaterra, Spain}

\date{\today}

\begin{abstract}
First-principles calculations of substitutional defects and vacancies are performed for
zigzag-edged hybrid C/BN nanosheets and nanotubes which recently have been proposed to 
exhibit half-metallic properties. The formation energies show that defects form 
preferentially at the interfaces between graphene and BN domains rather
than in the middle of these domains, and that substitutional defects dominate
over vacancies.  Chemical control can be used to favor localization 
of defects at C-B interfaces (nitrogen-rich environment) or C-N interfaces 
(nitrogen-poor environment). Although large defect concentrations have been considered 
here (10$^6$ cm$^{-1}$), half-metallic properties can subsist when defects are localized 
at the C-B interface and for negatively charged defects localized at the C-N interface, 
hence the promising magnetic properties theoretically predicted for these zigzag-edged
nanointerfaces might not be destroyed by point defects if these are conveniently 
engineered during synthesis.
\end{abstract}


\maketitle

\section{Introduction}

Hybrid C and BN nanostructures have been studied since the first attempts to dope carbon 
nanotubes with boron or nitrogen.~\cite{Stephan94} Nanotubes (NT) with C, B, and N were 
theoretically predicted\cite{Miyamoto94} and experimentally synthesized in the mid 
90s,~\cite{Stephan94,Chopra95,Redlich96,Suenaga97} but the recent advent of graphene 
has revitalized the field. Hybrid C and BN nanosheets are becoming accesible\cite{Han2008,Ci2010} 
offering a new route to enhance graphene's promise, enabling for fine-tuning the electronic 
and optical properties (for example modulation of the electronic bandgap, chemical reactivity, 
etc).

Following the advances in experimental growth of hybrid C-BN nanostructures,~\cite{Ci2010,
Enouz2007,Wei2011} the stability of domain-separated C and BN nanosheets and nanotubes has 
been studied by first principles simulations in the last couple of years.~\cite{Ding2009,
Du2009,Zhang2009,Ivanovskaya2009,An2010} There is general agreement that segregation is 
energetically advantageous, because B-N and C-C bonds are more stable than C-N and C-B and 
the former are favored by the formation of C and BN domains.  Furthermore, according to 
molecular dynamics simulations, hybrid C-BN armchair nanotubes can be spontaneously formed 
via the connection of zigzag-edged BN and graphene nanoribbons (BNNR and GNR) at room 
temperature,~\cite{Du2009} and their stability can be competitive with that of the corresponding 
pristine BNNTs and CNTs.~\cite{An2010} These zigzag-edged heterostructures are particularly 
appealing for spintronic applications, as the polarity of the BNNR gives an interfacial dipole, 
hence an effective electric field acting on the graphene ribbon.~\cite{Pruneda2010} It has been 
predicted that half-metallicity can be induced in zigzag-GNRs by application of a sufficiently 
large in-plane electric field perpendicular to the edges of the ribbon.~\cite{Son2006}  

Magnetism in zigzag-edged graphene-based nanostructures is related to the presence of 
electronic states that are mainly localized at the edges.~\cite{Fujita1996} These states 
seem to be ubiquitous in hexagonal structures with zigzag borders\cite{Okada2000}, 
and are not present in armchair edges. Roughness and defects have been shown to strongly 
affect the electronic properties of zigzag-GNR, inducing a continous decrease of the 
magnetic moments with increasing concentration of defects until the system becomes 
nonmagnetic for concentrations of one defect per $\sim$10\AA, which might be a typical 
concentration for real samples.~\cite{Huang2008} Nevertheless half-metallicity induced by 
external electric fields seems to survive at similar defect concentrations with the same 
critical field strength.~\cite{Son2006}  At zigzag interfaces between C and BN domains, 
however, defects can change the electronic screening of the edge polarity hence affecting 
the effective electric potential on graphene's ribbon and destroying its intrinsic 
half-metallicity. It is then important to address the role played by point defects at the 
boundary between graphene and BN domains.  In this work, first principles simulations 
are performed to determine the formation energies and electronic properties of intrinsic 
point defects (vacancies and substitutional defects) at zigzag interfaces between 
C and BN nanodomains that could exhibit magnetism.  It is shown that these defects are 
energetically favored when localized close to domain interfaces, and their electronic 
properties are determined by the presence of a defect-induced electronic state localized 
at the C-N interface which could be occupied or empty depending on the donor/acceptor 
nature of the defect. 

The paper is organized as follows. After describing the technicalities in section 
\ref{Method}, we present and discuss the findings in \ref{Results}. 
Results of the energetics of neutral defects at the interfaces are examined and 
compared to similar defects placed at a graphene-like or BN-like monodomain. 
Magnetism at the interfaces and other electronic properties are then analysed, and
the effect of defect concentration and charge state discussed. Finally, section
\ref{Conclude} summarizes the conclusions.

\section{Methodology}
\label{Method}
Ab initio pseudopotential density functional calculations are performed for both
planar (2D) superlattices\cite{Pruneda2010}, and (n,n) armchair nanotubes 
(5$\le$n$\le$14) with zigzag edged C-BN domains along the tube axis following 
the geometries discussed in the literature.~\cite{Du2009,Zhang2009,An2010,Huang2010}
Troullier-Martin type pseudopotentials\cite{pseudos} and numerical atomic orbitals 
with double-$\zeta$ plus polarization are used to describe the electronic valence 
states within the spin-polarized generalized-gradient approximation\cite{PBE} 
as implemented in the SIESTA code.~\cite{siesta}  The atomic 
positions are determined with a structural relaxation until the forces are smaller 
than 0.02eV/\AA. An accurate description of the boundary electronic states requires
a smooth sampling of the reciprocal space, and typically Monkhorst-Pack grids of 
at least 1$\times$1$\times$100 is used to sample the Brillouin zone. 

\begin{figure}
\includegraphics[width=0.45\textwidth]{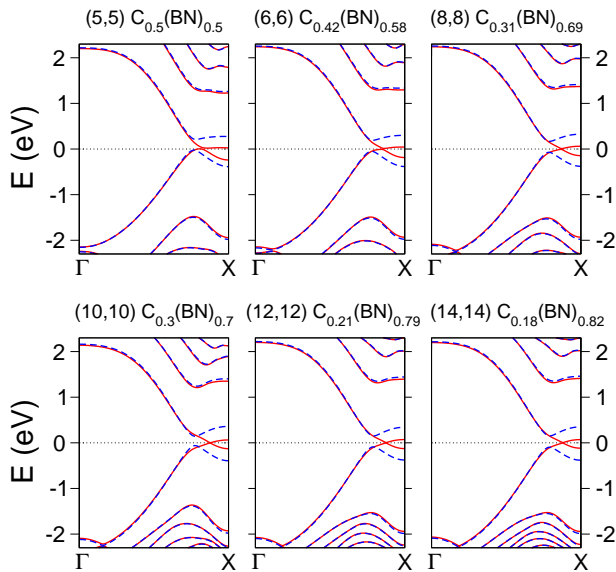}
\caption{\label{chiralities} (color-online) Bandstructures for hybrid C-BN 
nanotubes with different armchair chiralities and zigzag edges between C- and 
BN-domains along the tube axis. Different relative compositions of C$_x$(BN)$_{1-x}$ 
are picked to ensure half-metallicity.  The absence/presence of a band gap for up/down 
spin polarizations is revealed in the solid (red) or dashed (blue) lines.}
\end{figure}

It has been claimed that half-metallicity in C-BN nanotubes can be achieved within 
a certain ratio of carbon content (30\%-50\%) and is not dependent on the diameter 
of the tube.~\cite{Huang2010} This allegation can be misleading. Although half-metallic 
ground states can be obtained for different chiralities (see Fig.\ref{chiralities}), 
this only happens for C-concentrations under a critical value, that does depend on the 
tube diameter.  Analogously to the case of planar C-BN heterostructures, the critical 
parameter is the relative width of the C- and BN-strips.~\cite{Pruneda2010} To address 
the stability of magnetism to the presence of defects, systems that hold half-metallic
properties in their pristine (defect-free) structure, are considered. 
In particular, throughout this work we take: (i) nanotube with chirality (5,5) 
and 50\% C concentration (corresponding to 5 C and 5 BN zigzag chains), 
and (ii) planar (5,7) C-BN superlattice (composed by 5 C and 7 BN zigzag chains).
Native point defects are introduced near the edges of the C and BN domains in a supercell 
geometry, with at least 30 \AA\ empty space in-between periodic images.  
Different concentrations can be modeled by changing the size of the supercell along 
the periodic axis.  If not otherwise stated, supercells made out of 8 repetitions of 
the unit cell along the tube axis are considered (a total of 160 atoms). 

The formation energy of a defect in charge state $q$ is defined in terms of the 
chemical potentials of the species involved in the defect, and the Fermi level, 
$\mu_e$, measured relative to the top of the valence band $E_v$:
\begin{equation}
\label{Eformation}
E_f(X)=E_{tot}(X)-\sum_i{n_i\mu_i} + q(\mu_e+E_v)  \nonumber
\end{equation}
where $E_{tot}$ is the energy of the defective supercell containing $n_i$ atoms of 
species $i$ (C, B, and N) with chemical potential $\mu_i$, at $T=0$ (entropic contributions 
are neglected).  These chemical potentials are specified by a reference system that acts 
as the reservoir of atoms. Typically reference values used in the literature are taken 
from graphene, molecular nitrogen, and bulk boron. These, however, will only give formation 
energies under limits of high concentrations of C, N, or B. Formally, one can use the formation 
energies of these reference systems as thermodyncamic limits to the chemical potentials, 
which must satisfy the following conditions:
\begin{enumerate}
\item stability of the C-BN complex:
\begin{equation}
\label{chempot}
\mu_C+\mu_B+\mu_N=\Delta E_f(\text{CBN})
\end{equation}
\item The values that cause precipitation into its constituents,
\begin{eqnarray}
\nonumber
\mu_C\le0, \hspace{0.3cm} \mu_B\le0,\hspace{0.3cm} \mu_N\le0 \nonumber
\end{eqnarray}
\item The values that cause formation of each domain
\begin{eqnarray}
\nonumber
\mu_B+\mu_N \le \Delta E_f(\text{BN}) \\
\nonumber
\mu_C \le \Delta E_f(\text{C}) 
\end{eqnarray}
\end{enumerate}
where $\Delta E_f$ denotes the generalized formation free energy of the corresponding 
system relative to pure C or BN nanotube. Relation (\ref{chempot}) can be used to write 
one of the atomic chemical potentials in terms of the other two. Considering that C-doping 
in BNNT had proved elusive\cite{C-doping}, whereas BNNT can be synthesized by substitutional 
reaction from CNT\cite{Han1998}, in the following the chemical potential of C will be set to 
the value that corresponds to clean CNT, meaning that the defect formation energies will be 
given in the C-rich limit, and only $\mu_N$ will be used as free parameter.

\section{Results and Discussion}
\label{Results}

\begin{figure}
\includegraphics[width=0.45\textwidth]{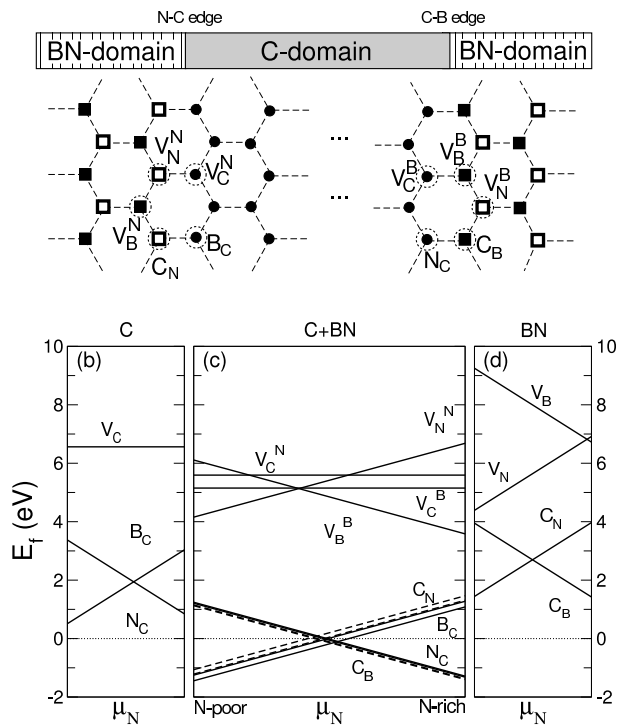}
\caption{\label{Energies} (a) Atomic positions considered for the different defects. 
Circles, filled squares, and empty squares denote C, B and N atoms. Periodic boundary 
conditions require that both C-N and C-B edges are included in the simulation box. 
Formation energies (in eV) for the studied neutral defects as a function of Nitrogen 
chemical potential ($\mu_N$) are shown in the lower panel for: (b) the pure CNT, 
(c) hybrid C-BN nanotubes (solid lines) with 8 repetitions of the unit cell, 
and 2D nanosheet (dashed lines) with 7 repetitions of the unit cell, and 
(d) BN nanotubes with (5,5) chirality.}
\end{figure}

Substitutional carbons on the boron and nitrogen sublattices (C$_{\text B}$ and C$_{\text N}$), 
boron or nitrogen atoms in carbon sites (B$_{\text C}$ and N$_{\text C}$) and vacancies 
of each species (V$_{\text C}$, V$_{\text B}$ and V$_{\text N}$) were considered, as shown
in figure \ref{Energies}. 
Ghost orbitals are used to improve the localized atomic orbital description at the 
vacancy sites.  Labels $N$ and $B$ will be used to denote the edge at which the defect 
is placed, where $N$ designates the interface with $C-N$ bonds, whereas $B$ is the edge 
with $C-B$ bonds. Accordingly defects are classified into {\it set N} 
(C$_{\text N}$, B$_{\text C}$,  V$_{\text N}^N$, V$_{\text B}^N$, or V$_{\text C}^N$), 
and {\it set B} (C$_{\text B}$, N$_{\text C}$,  V$_{\text N}^B$, V$_{\text B}^B$, or
V$_{\text C}^B$). 

\begin{figure*}
\includegraphics[width=0.75\textwidth]{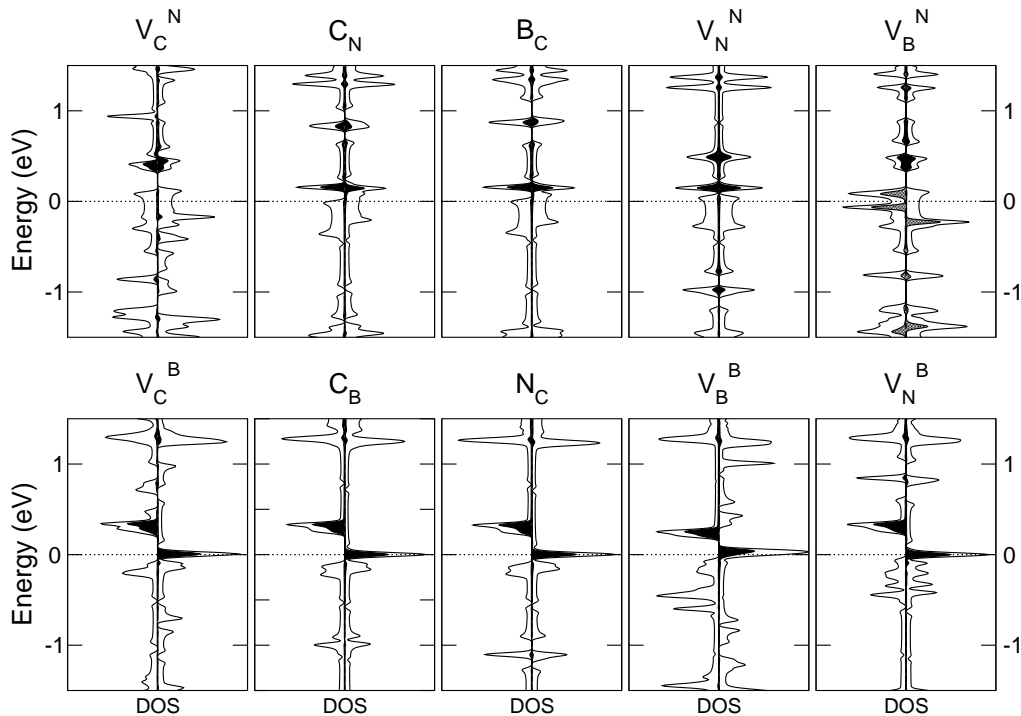}
\caption{\label{pdos} Density of States (DOS) around the Fermi level (at 0 eV) 
for defects localized at the C-N (top) and C-B (bottom) interfaces. 
Dark shaded regions correspond to the projected DOS (pDOS) on C orbitals localized 
at the N-interface.
Gray regions for  V$_{\text C}^{\text N}$ and V$_{\text B}^{\text N}$ are the pDOS 
on C or N orbitals close to the C or B vacancy, respectively.}
\end{figure*}

The formation energies for the interfacial (neutral) defects in (5,5) C-BN nanotube are 
shown in Fig.\ref{Energies} and compared to the energetics of the same defects in a pristine 
pure-CNT and pure-BNNT of the same chirality.  The energies of pure C and BN nanotubes 
are in agreement with values published in the literature\cite{Krasheninnikov2006,
Koretsune2008,Piquini2005}, considering that the formation energy depends on the chirality, 
and increases with tube's diameter\cite{Krasheninnikov2006,Gou2007}. 
The cost of removing atoms (vacancy formation) is generally larger than the energy of 
atom substitution (disorder), which is indicative of the high stability of the 
honeycomb structure.
First principles calculations have shown that vacancy formation is energetically more favorable 
near the edge of zigzag BNNRs than in the center of the ribbon\cite{Tang2010}. A similar effect 
is observed now, with defects at the edge being at least 1 eV less energetic. Interestingly, 
recent experiments have shown that carbon substitution (doping) in BN nanostructures is 
favored at the edges.~\cite{Wei2011}  

Note that C$_{\text N}$ and B$_{\text C}$ (defects of {\it set N}) are energetically 
favorable in a nitrogen-poor environment (B-rich), and the same is true for N$_{\text C}$ 
and C$_{\text B}$ ({\it set B}) in N-rich situation, so that the localization of defects 
at each edge of the heterojunction could be chemically controled. Furthermore, when neutral 
C$_{\text N}$ and B$_{\text C}$ defects are formed, one hole is added to the system, so that 
these defects can act as electron traps. On the other hand, N$_{\text C}$ and C$_{\text B}$ 
add one extra electron to the system and could possibly become positively charged under 
appropriate conditions (low electron chemical potential $\mu_e$).  In pure BN nanosheets 
and nanotubes, V$_{\text B}$ is also an acceptor that can trap charges to form 
V$_{\text B}^{-}$, whereas V$_{\text N}$ has two one-electron states in the gap, 
one occupied and the other empty, enabling the formation of both V$_{\text N}^{+}$ 
or V$_{\text N}^{-}$, depending on the value of $\mu_e$.  


Figure \ref{pdos} shows the electronic Density of States (DOS) for {\it set N} (top) and 
{\it set B} defects.  Notice that, even for the high defect concentration considered in 
these simulations (5$\cdot 10^6$cm$^{-1}$), the later remain half-metallic, with a sharp peak
in the DOS at the Fermi level. This peak corresponds to a defect-induced electronic state that
{\it is localized at the N-edge rather than at the defect itself} (placed at the B-edge). 
Hence, electrons coming from these {\it donor defects} are transferred through the 
C-domain to the region with lower electronic potential, as described in reference 
\onlinecite{Pruneda2010}.  This can be seen from the DOS projected on the carbon 
orbitals localized at the C-N interface (shaded regions in the figure). 
This localized band has some dispersion along the edge axis (there is slightly more 
contribution to the pDOS from C atoms directly opposite to the defect site), and
a larger unit cell would be needed to address its localization length. 
The exchange splitting for this band is $\sim$0.3 eV for all the {\it set B} defects, 
and the calculations give a total magnetic moment of $\sim$1 $\mu_B$ for C$_{\text B}$, 
N$_{\text B}$ and V$_{\text N}^{\text B}$, 2 $\mu_B$ for V$_{\text C}^{\text B}$, 
and $\sim$3 $\mu_B$ for V$_{\text B}^{\text B}$.  

On the other hand, the missing electron (extra hole) for {\it set N} defects has a dramatic 
effect the corresponding DOS, that does not show half-metallic properties. There is a 
similar defect-induced electronic state localized at the C-N interface for C$_{\text N}$,
B$_{\text C}$ and V$_{\text N}^{\text N}$, but now it is unoccupied ($\sim$0.1 eV 
above the Fermi level) and not spin-polarized (C$_{\text N}$ and B$_{\text C}$ have a 
total magnetic moment of $\sim$0.5 $\mu_B$).
V$_{\text C}^{\text N}$ and V$_{\text B}^{\text N}$ are slightly different, with an empty 
C-N interfacial state at $\sim$0.4 eV above the Fermi level, and a rather localized electronic
state on the atoms surrounding the vacancy site (shaded light regions in the figure) which
is spin-polarized (total magnetization of $\sim$1.8 $\mu_B$ and 1$\mu_B$, respectively.
The lowering of the Fermi level means a partial depopulation of the $\pi_{\text B}$ 
state,~\cite{Pruneda2010} and a weakening of the magnetic instability induced by 
electron-electron interactions\cite{Fujita1996}. The reduction of the electronic 
occupation of the $\pi_{\text B}$ state lowers the electronic potential at the B-edge, 
and significant decrease of the {\it induced} electric field in the graphene 
ribbon, so that half-metallicity cannot survive.~\cite{Pruneda2010} 

\begin{figure}
\includegraphics[width=0.45\textwidth]{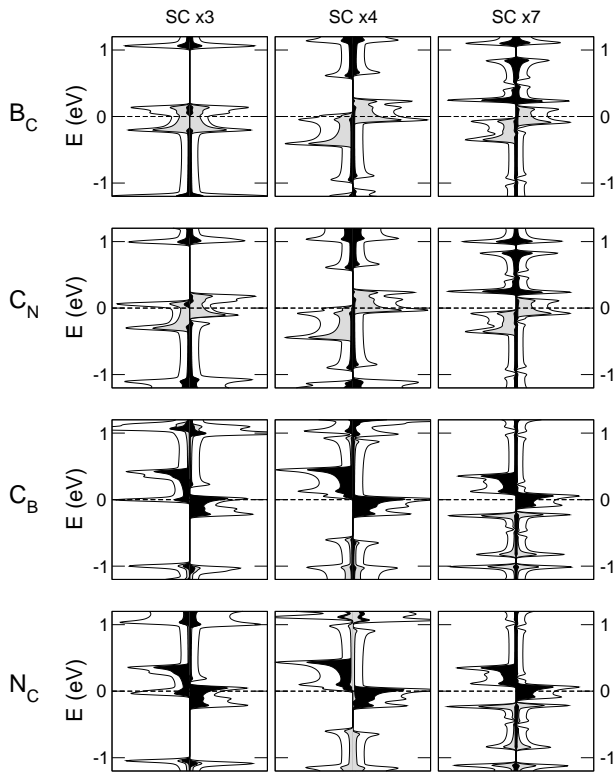}
\caption{\label{concentration} Density of States for substitutional defects as a 
function of defect concentration for planar C/BN superlattices (similar results 
are obtained for nanotube geometries). From left to right, supercells with 3, 4 
and 7 units of repetition along the interfacial axis, corresponding to concentrations of
13, 10 and 6$\cdot$10$^6$ cm$^{-1}$. Dark/light shaded regions 
correspond to projected DOS over C orbitals at the N/B interface.}
\end{figure}

Let's focus in the following on those defects that have the lower formation energies, 
i.e. substitutionals.  The effect of increasing the defect concentration even further 
is shown in figure \ref{concentration}, where the DOS of different supercells of planar
bidimensional superlattices are shown.  Small differences arise between this planar 
geometry and the previous tubular case, mainly because of the different C/BN ratio, 
but the general results are equivalent.  For p-type {\it set N} defects (B$_{\text C}$ 
and C$_{\text N}$) the reduced number of electrons shifts the Fermi level towards lower 
values as we move towards higher defect concentrations (smaller supercells, from 
right to left), depopulating the initially fully occupied majority spin band, localized 
at the C-B interface (light shaded region), until the full spin polarization at the 
Fermi level is destroyed. On the other hand, for n-type {\it set B} defects 
(N$_{\text C}$ and C$_{\text B}$), the extra electrons added to the system shifts 
the Fermi level to higher values as we move towards higher concentrations, increasing 
the occupation of the C-N electronic states, and eventually destroying half-metallicity.
Notice again that defects localized at the C-B interface add extra electrons that
are transferred to the C-N interface, while defects localized at the C-N interface, 
which are electron deficient tend to remove charge from the C-B interface.

For the limit of low concentration of defects, the semimetallic graphene domain 
could be considered as the electron reservoir so that $\mu_e$ is fixed by the position 
of GNR's Fermi level, and hence only neutral defect calculations were considered up 
to now.  However the donor/acceptor character of the defects has become relevant when 
discussing their electronic properties, revealing the need to address charged defects.
When addressing charge states for defects in semiconductors and insulatores, the value
of $\mu_e$ in equation (\ref{Eformation}) ranges between zero 
(the top of the valence band) and the energy band gap, and a careful alignment of the
reference energy has to be considered to compare energetics of neutral and charged defects.
Here, however, the system is semimetallic regardless of the charge state, so that the
position of the Fermi level can be used to align the energetics, and $\mu_e=0$ can be
used to compare formation energies.

\begin{figure}[b]
\includegraphics[width=0.45\textwidth]{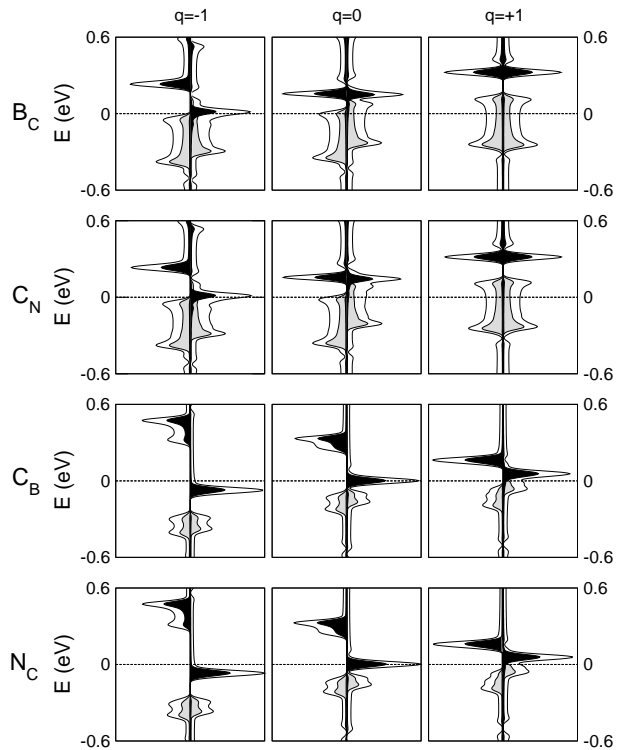}
\caption{\label{charged} Density of States for charged ($q=-1, 0, +1$) substitutionals. 
From top to bottom B$_{\text C}$, C$_{\text N}$ ({\it set N}), C$_{\text B}$, 
and N$_{\text C}$ ({\it set B}). Dark (light) shaded regions correspond to the pDOS on C orbitals
localized at the N-interface (B-interface).}
\end{figure}

As shown in figure \ref{charged}, when an extra electron is added to the system 
($q=-1$), it localizes at the N-edge state (dark peaks) close to the Fermi level, 
recovering half-metallicity for the {\it set N} defects (the states become fully 
occupied for {\it set B} defects which remain half-metallic). The formation energies 
for these charged defects are reduced by $\sim$2.3 eV (at $\mu_e=0$ eV).  There is 
a decrease in the system's total magnetization for {\it set N} ($\sim$0.1 $\mu_B$) 
and an increase to 2 $\mu_B$ for {\it set B}. On the other hand, removal of an 
electron (giving defects with charge state $q=+1$) has negative effects on the 
electronic properties of substitutionals: the lowering of the Fermi level 
depopulates the N-edge state for {\it set B} defects (although the total DOS 
remains half-metallic), and destroys magnetization for all defects ({\it set N} 
remains metallic). In this case, the formation energies are increased by 
$\sim$3.6 eV, clearly destabilizing these positively charged defects.

Following the above results, a N-rich environment together with a high $\mu_e$ 
(providing electrons to the system) would increase the possibilities for subsistence 
of half-metallicity in these hybrid C-BN structures. B-edge localized defects are
favored by an N-rich atmosphere and even at large defect concentrations 
(10$^{6}$ cm$^{-1}$) conducting electrons remain fully spin-polarized. On the other 
hand, under these chemical conditions, there would be a lower concentration of defects 
localized at the N-edge (higher formation energies), which although hinder
half-metallicity when neutral, they do not in their negative charge state. 
It is interesting that similar conditions have been suggested to be best for 
manufacturing hybrid compounds through electron-beam-induced doping of BN 
nanostructures with C.~\cite{Berseneva2011}

\section{Conclusions}
\label{Conclude}
First principles DFT calculations have shown that intrinsic defects in hybrid 
C-BN nanostructures are more stable at the boundaries between C and BN domains.
Both bidimensional superlattices of alternating graphene and BN nanoribbons, and
armchair nanotubes with axial domains of C and BN that have zigzag edges have been
analysed, giving similar results. In these systems, the electronic bandstructure close 
to the Fermi level is determined by characteristic zigzag-edge states localized at the 
boundary of the hexagonal lattice, as already discussed in the literature.

The stability of the honeycomb structure results in a higher formation energy 
for vacancies (V$_{\text C}$, V$_{\text B}$, or V$_{\text N}$) than for 
substitutionals. 
While a nitrogen-poor environment favors the formation of defects at C-N interfaces 
(mainly C$_{\text N}$ and B$_{\text C}$), nitrogen-rich atmospheres favors defects 
localized at the C-B boundary (C$_{\text B}$ and N$_{\text C}$), which could be
classified as electron acceptors and donors respectively. Extra electrons provided
by the latter are transferred to a defect-induced partially occupied electronic 
state localized at the opposite edge (C-N) and the system would remain
half-metallic, even for the large defect concentrations considered here. 
If extra electrons are added to the system, they are transferred to this 
electronic state and the defect would be further stabilized (in their negatively 
charged state), whereas electron-deficiency drains the electronic level and increases
the defect formation energy. On the other hand, substitutional defects localized at 
the C-N interface, add one hole to the system hence lowering the Fermi level and
weakening the magnetic instability (through depopulation of the $\pi_{\text C-B}$ orbital)
resulting in the cancelation of half-metallicity. The defect-induced electronic state
localized at the C-N interface is still there, but unoccupied. Extra electrons added to 
the system would populate this level, recovering half-metallicity and lowering the 
defect formation by a couple of eV.

Finally, the analysis of the magnetic properties reveals that half-metallicity 
(i) could survive defect concentrations at the interfacial line of up to 
10$^{7}$ cm$^{-1}$, and (ii) could be tuned through appropriate electronic 
doping that avoids an excesive charge depletion from the C-B interface for 
high concentrations of  p-type substitutionals or excesive charge increase 
at the C-N interface for n-type substitutionals. Experimental investigation of
magnetism in these hybrid nanostructures is encouraged.

\begin{acknowledgments}
The author acknowledges financial support by the Spanish MCINN 
(FIS2009-12721-C04-01 and CSD2007-00041). 
\end{acknowledgments}


\bibliography{biblio}

\end{document}